%% file: ECAL-lhcp.tex
\documentclass[10pt]{article}
\usepackage{graphicx}

\def\Title#1{\begin{center} {\Large #1 } \end{center}}
\def\Author#1{\begin{center}{ \sc #1} \end{center}}
\def\Address#1{\begin{center}{ \it #1} \end{center}}

\newcommand\pubblock{\rightline{\begin{tabular}{l} Proceedings of the Second Annual LHCP\\ \pubnumber\\
         \pubdate  \end{tabular}}}

\newenvironment{Abstract}{\begin{quotation} \begin{center} 
             \large ABSTRACT \end{center}\bigskip 
      \begin{center}\begin{large}}{\end{large}\end{center} \end{quotation}}

\newenvironment{Presented}{\begin{quotation} \begin{center} 
             PRESENTED AT\end{center}\bigskip 
      \begin{center}\begin{large}}{\end{large}\end{center} \end{quotation}}

\input econfmacros.tex

\usepackage{color}

\newcommand\pubnumber{ }

\newcommand\pubdate{\today}

\def\affiliation{
On behalf of the CMS collaboration, \\
Northeastern University, Boston, 02115, U.S.A }

\begin{document}

\large
\begin{titlepage}
\pubblock

\vfill
\Title{  Evolution studies of the CMS ECAL endcap response and upgrade design options for High-Luminosity LHC  }
\vfill

\Author{ Andrea Massironi }
\Address{\affiliation}
\vfill
\begin{Abstract}

High-Luminosity running at the LHC, which is planned for 2022 and beyond,
will imply an order of magnitude increase in radiation levels and
particle fluences with respect to the present LHC running conditions.
The performance evolution of the CMS electromagnetic calorimeter (ECAL),
comprising 75,848 scintillating lead tungstate crystals,
indicates that an upgrade of its endcaps will be needed for HL-LHC running,
to ensure an adequate performance. Results from LHC collision periods,
beam tests and laboratory measurements of proton-irradiated crystals
are combined to predict the performance of the current detector at the HL-LHC.
In addition, an overview is given of various R\&D studies
towards a replacement of the ECAL endcaps for the HL-LHC running period.


\end{Abstract}
\vfill

\begin{Presented}
The Second Annual Conference\\
 on Large Hadron Collider Physics \\
Columbia University, New York, U.S.A \\ 
June 2-7, 2014
\end{Presented}
\vfill
\end{titlepage}
\def\thefootnote{\fnsymbol{footnote}}
\setcounter{footnote}{0}
%

\normalsize 


\section{Introduction}

Proton-proton collisions detected by the Compact Muon Solenoid (CMS) and the ATLAS
experiments at the Large Hadron Collider (LHC) have let to the discovery of a Higgs boson~\cite{Chatrchyan:2012ufa, Aad:2012tfa}.
The experimental evidence for the new particle is most striking in the Higgs decay modes into four leptons
and into two photons. The detection of these decay modes depends on very good photon 
and electron identification capabilities, energy resolution and response linearity.
The CMS detector was specifically optimized for these signatures
with the Electromagnetic Calorimeter (ECAL) being a central component of
the detector design:
the ECAL is a compact, hermetic, fine-grained and homogeneous calorimeter
made of 75848 lead tungstate (PbWO$_4$) scintillating crystals arranged in a quasi-projective
geometry and organized into a barrel region (EB), with pseudorapidity coverage up to $|\eta|=$~1.48,
closed by two endcaps (EE) that extend up to $|\eta|=$~3. The choice of PbWO$_4$ was driven
by its small radiation length (0.89 cm) and its small Moli\`ere radius (2.19 cm). These properties
allow a compact design, fitting into the CMS solenoid, and a lateral segmentation of 1 degree in
EB.
The fast response of PbWO$_4$ (99\% of the light is collected in 100 ns)
is compatible with the 40 MHz collision rate of the LHC. The
scintillation light of each crystal is read out in EB with two avalanche photodiodes (APDs)
per crystal
and in EE with one vacuum phototriode (VPT)
per crystal.

\section{Evolution studies}

\begin{figure}[hbtp]
\centering
\includegraphics[height=2.3in]{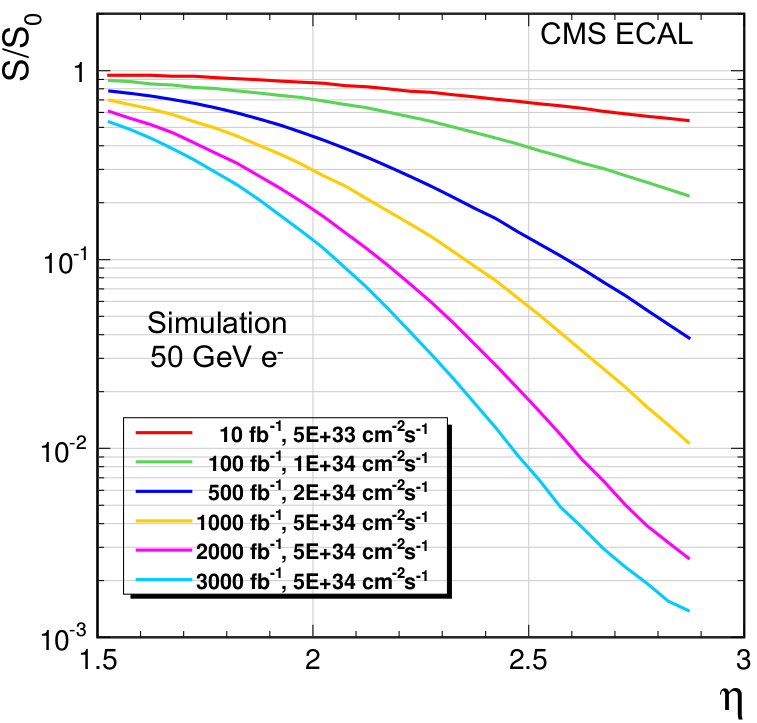}
\includegraphics[height=2.3in]{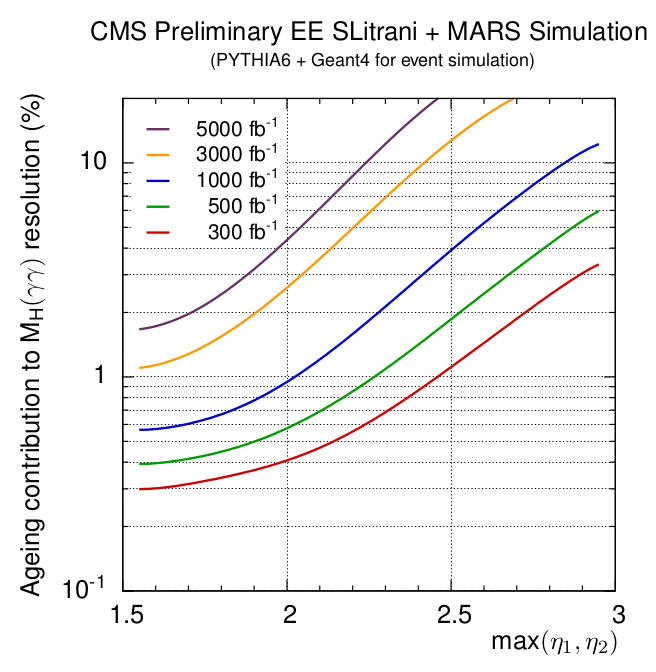}
\caption{
  The extrapolation of the response evolution of the ECAL endcap up to an
  integrated luminosity of 3 ab$^{-1}$ (left).
  Contribution to the Higgs mass resolution due to the ECAL ageing
  in EE (right), for the two photon decay of
  the Higgs boson, as expected as a function of the maximum pseudo-rapidity of the photon pair, for several
  conditions of integrated luminosities.
  A full simulation of the electromagnetic shower development in the crystals is performed with Geant4, 
  while the radiation damage is modeled using the MARS simulation~\cite{mars} and 
  the ray-tracing program SLitrani~\cite{slitrani} is used to model the light output.
 }
\label{fig:evolution}
\end{figure}

The energy resolution of the CMS ECAL can
be written as the quadratic sum of 3 contributions:
$$
\left(\frac{\sigma_{E}}{E} \right) =      \left( \frac{\sigma_{stat}}{\sqrt{E}}\right) \oplus \left( \frac{\sigma_{noise}}{E}\right)  \oplus c.
$$
The stochastic term ($\sigma_{stat}$), the electronic noise ($\sigma_{noise}$) and the constant term (c), that were measured
in the test beam, have been shown to match the design requirements: 2.8\%, 120 MeV and 0.3\%
respectively in the barrel for energy reconstruction based on a 3x3 crystal matrix~\cite{jinst}.
The ECAL was initially designed to operate for 10 years of LHC running, at a peak luminosity of
L=10$^{34}$cm$^{-2}$s$^{-1}$, up to a total integrated luminosity of 500 fb$^{-1}$.
In order to fully exploit the LHC potential, a major upgrade is foreseen in 2022: 
the High-Luminosity LHC (HL-LHC), a new machine designed for high precision measurements and possible new
discoveries, is expected to operate for 10 years and to collect 3 ab$^{-1}$ by the end of 2033.
During this Phase II (HL-LHC) operation, there will be a very challenging running environment
with a peak luminosity 5 times the initial LHC design conditions and radiation levels typically a factor
6 higher, with strong pseudo-rapidity dependence in the EE. 
The HL-LHC represents a great opportunity to explore a
rich physics program, however, the detectors must be optimised in order to fulfill the requirements
for the physics plans.\\
During current operation, the ECAL has experienced some radiation damage: by monitoring the
evolution of the ECAL performance, it is possible to model such effects and provide long term
projections for future runs~\cite{Chatrchyan:2013dga}.
While the transparency loss due to 
electromagnetic damage
is recoverable,
as it has been seen during non-collision periods of the 7 and 8 TeV run, 
the hadron-induced damage is irrecoverable:
it has been shown how PbWO$_4$ exposed to hadronic
showers from high-energy protons and pions experiences
a cumulative loss of light transmission which is permanent at room temperature~\cite{jinst2,Lucchini}.
Figure~\ref{fig:evolution} (left) shows the extrapolation of the
response evolution of the ECAL endcap up to an
integrated luminosity of 3 ab$^{-1}$.
A loss of the ECAL response is translated into a degradation of the detector performances,
as shown in 
Figure~\ref{fig:evolution} (right),
where the contribution to the Higgs mass resolution due to the ECAL ageing is reported.

\section{Possible design options for electromagnetic calorimetry at the HL-LHC}

All the options for the upgrade of the current calorimeter must be able to operate in the extremely
hostile environment of the HL-LHC with good performance. All the components of the upgraded
detector, such as scintillators, wavelength shifters (WLS), photo-detectors and electronics will have
to be sufficiently radiation-tolerant to withstand the ionising doses.
Also the Hadronic Calorimeter (HCAL) Endcap needs to be replaced for the HL-LHC operation.
Two different upgrade scenarios are being considered: 
in the first one, the new ECAL endcap is designed in a standalone configuration,
while in the second scenario the replacement of both the forward ECAL and HCAL
with a common integrated calorimeter is being considered.\\
Although the EE requires replacing, the EB has been designed to endure up to 3 ab$^{-1}$ and
maintain its performances.

\section{Scenario 1: Shashlik}

\begin{figure}[hbtp]
\centering
\includegraphics[height=1.4in]{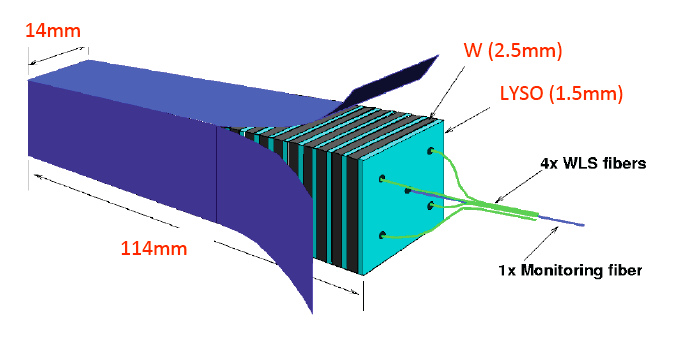}
\includegraphics[height=1.4in]{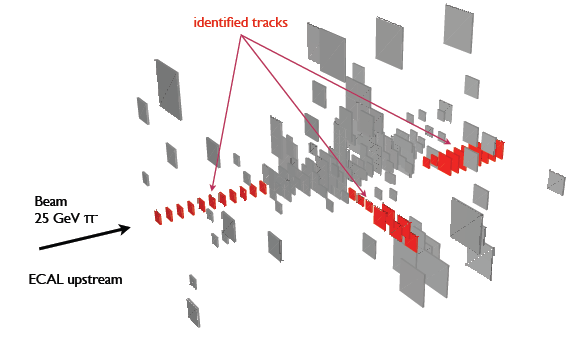}
\caption{
  Conceptual layout
  of a LYSO-Tungsten sampling
  calorimeter (left), read out by wavelength
  shifting fibers. This option achieves
  an energy resolution of about 1\%
  for photons  
  with energy greater than 100 GeV
  according to GEANT4 simulations.  
  A pion shower being reconstructed in a conceptual HGC (right).
 }
\label{fig:Shashlik}
\end{figure}

A sampling calorimeter has been proposed, using radiation-hard inorganic scintillators such as LYSO or
CeF$_3$, in order to reduce the effects of radiation damage.
Detailed simulations have been carried out with a sampling configuration read out by wavelength shifting fibers.
Standalone GEANT4
simulations have demonstrated that the desired performance could be achieved with a configuration
using 29 layers of LYSO plates of 2 mm thickness, interleaved with 28 W absorber plates, resulting
in a total cell depth of 170 mm.
The current EE should be replaced by 60800 Shashlik modules, whose design
is shown in Figure~\ref{fig:Shashlik}.

\section{Scenario 2: HGC, High Granularity Calorimeter}

A high granularity calorimeter with a detailed sampling in both the hadronic and the electromagnetic
sections with excellent pointing capability is being considered:
layers of silicon detectors are alternated with samplings of lead or brass, that feature very
high longitudinal and lateral granularities in the electromagnetic and the front hadronic calorimeter
sections,
and a coarser segmentation backing hadronic calorimeter section.
It is being optimised for particle flow algorithms, to separate and track showers as they develop.
A sketch of pion shower being reconstructed
in a conceptual high granularity calorimeter is shown in Figure~\ref{fig:Shashlik} (right).
%
%

\section{Conclusions}

Although the LHC is at the beginning of its operation, with about 30 fb$^{-1}$ of integrated luminosity,
evidence of some radiation damage is already visible in the ECAL. The observation is in general
agreement with expectations and is taken into account in data analysis to ensure the high quality
of the ECAL detector performance and the corresponding physics results.
In ten years from now, the HL-LHC will operate at a factor 5 higher instantaneous luminosity with
respect to LHC, eventually delivering up to 3 ab$^{-1}$ in Phase II. Such challenging conditions 
impose stringent detector requirements in terms of performance and radiation-hardness. Simulation
studies have been carried out, to predict the evolution of the ECAL response in the high-radiation
environment and to understand the requirements for the detector upgrade in order to maintain a
good level of performance.
Several R\&D studies have investigated the best upgrade options for the forward region of
the CMS calorimeter, in order to exploit the physics potential offered by the HL-LHC,
while the EB has been proven to maintain high performances with the current setup 
up to 3 ab$^{-1}$.

%


%
%
%
%


\end{document}

%% file: econfmacros.tex
\def\beq{\begin{equation}}
\def\eeq#1{\label{#1}\end{equation}}
\def\eeqn{\end{equation}}

\def\beqa{\begin{eqnarray}}
\def\eeqa#1{\label{#1}\end{eqnarray}}
\def\eeqan{\end{eqnarray}}

\let\bar=\overbar

\def\Dslash{\not{\hbox{\kern-4pt $D$}}}
\def\dslash{\not{\hbox{\kern-2pt $\del$}}}

\def\msb{{\bar{\ssstyle M \kern -1pt S}}}

